\documentclass{ws-ijmpa}

\begin{document}

\markboth{G.~Mao, J.~Gong, H.~St\"{o}cker, and W.~Greiner}{Energy
Spectra of Anti-nucleons in Finite Nuclei}

\catchline{}{}{}{}{}

\title{ENERGY SPECTRA OF ANTI-NUCLEONS IN FINITE NUCLEI}

\author{G.~MAO$^{1,2,3}$ and J.~GONG$^{1}$}
\address{$^{1}$Institute of High Energy Physics, Chinese Academy of Science \\
      P.O. Box 918(4), Beijing 100049, P.R. China\\
  $^{2}$Institute of Theoretical Physics, Chinese Academy of Science\\
      P.O. Box 2735, Beijing 100080, P.R. China \\
  $^{3}$CCAST (World Lab.), P.O. Box 8730, Beijing 100080, P.R.
       China }

\author{H.~ST\"{O}CKER and W.~GREINER}
\address{Institut f\"{u}r Theoretische Physik der J.W. Goethe-Universit\"{a}t\\
   Postfach 11 19 32, D-60054 Frankfurt am Main}
\maketitle
\begin{abstract}
The quantum vacuum in a many-body system of finite nuclei has been
investigated within the relativistic Hartree approach which
describes the bound states of nucleons and anti-nucleons
consistently. The contributions of the Dirac sea to the source
terms of the meson-field equations are taken into account up to
the one-nucleon loop and one-meson loop. The tensor couplings for
the $\omega$- and $\rho$-meson are included in the model. The
overall nucleon spectra of shell-model states are in agreement
with the data. The calculated anti-nucleon spectra in the vacuum
differ about 20 -- 30 MeV with and without the tensor-coupling
effects.
\end{abstract}
\vspace{0.5cm} \noindent It is quite interesting to study the
structure of quantum vacuum in a many-body system, e.g., in a
finite nucleus where the Fermi sea is filled with the valence
nucleons while the Dirac sea is full of the virtual
nucleon--anti-nucleon pairs \cite{Aue86,Rei86}. The shell-model
states have been theoretically and experimentally well established
\cite{Boh69}. It is the aim of our work to investigate the bound
states of anti-nucleons in the Dirac Sea. The observation of
anti-nucleon bound states is a verification for the application of
the relativistic quantum field theory  to a many-body system
\cite{Ser86}. Since the bound states of nucleons are subject to
the cancellation of scalar and vector potentials $S+V$ ($V$ is
positive, $S$ is negative) while the bound states of anti-nucleons
are sensitive to the sum of them $S-V$, consistent studies of both
the nucleon and the anti-nucleon bound states can determine the
individual $S$ and $V$. In addition, the knowledge of potential
depth for anti-nucleons in the medium is a prerequisite for the
study of anti-matter and anti-nuclei in relativistic heavy-ion
collisions.

We have developed a relativistic Hartree approach which describes
the bound states of nucleons and anti-nucleons in a unified
framework \cite{Mao99,Mao03}. In finite nuclei the Dirac equation
of nucleons is written as
 \begin{eqnarray}
 i \frac{\partial}{\partial t}\psi({\bf x},t) &=&
        \left[ -i \mbox{\boldmath $\alpha$} \cdot \mbox{\boldmath $\nabla$}
        + \beta\left(M_{N} - {\rm g}_{\sigma}\sigma({\bf x})\right)
        +{\rm g}_{\omega}\omega_{0}({\bf x})
        -\frac{f_{\omega}}{2M_{N}}i\mbox{\boldmath $\gamma$}\cdot
        \left(\mbox{\boldmath $\nabla$} \omega_{0}({\bf x})\right)
        \right. \nonumber \\
    &+& \left. \frac{1}{2}{\rm g}_{\rho}\tau_{0}R_{0,0}({\bf x})
        -\frac{f_{\rho}}{4M_{N}}i\tau_{0}\mbox{\boldmath $\gamma$}\cdot
        \left(\mbox{\boldmath $\nabla$}R_{0,0}({\bf x})\right)
        +\frac{1}{2}e(1+\tau_{0})A_{0}({\bf x})\right] \psi({\bf x},t).
        \label{field}
 \end{eqnarray}
The field operator can be expanded according to nucleons and
anti-nucleons and reads as
 \begin{equation}
\psi({\bf x},t)=\sum_{\alpha} \left[ b_{\alpha}\psi_{\alpha}({\bf
 x}) e^{-i E_{\alpha}t} + d^{+}_{\alpha}\psi^{a}_{\alpha}({\bf x})
 e^{i \bar{E} _{\alpha} t} \right]. \label{opera}
  \end{equation}
Here the label $\alpha$ denotes the full set of single-particle
quantum numbers.  The wave functions of nucleons and anti-nucleons
can be specified as \cite{Bjo64,Jin88}
  \begin{equation}
 \psi_{\alpha}({\bf x})= \left( \begin{array}{l}
 i \frac{G_{\alpha}(r)}{r} \Omega_{jlm}(\frac{{\bf r}}{r})  \\
  \frac{F_{\alpha}(r)}{r}\frac{\mbox{\boldmath $\sigma$}\cdot {\bf r}}{r}
 \Omega_{jlm}(\frac{{\bf r}}{r}) \end{array} \right), \label{wfn}
  \end{equation}
  \begin{equation}
 \psi_{\alpha}^{a}({\bf x})= \left( \begin{array}{l}
- \frac{\bar{F}_{\alpha}(r)}{r}\frac{\mbox{\boldmath
$\sigma$}\cdot {\bf r}}{r}
 \Omega_{jlm}(\frac{{\bf r}}{r})  \\
  i \frac{\bar{G}_{\alpha}(r)}{r} \Omega_{jlm}(\frac{{\bf r}}{r})
 \end{array} \right), \label{wfan}
  \end{equation}
where $\Omega_{jlm}$ are the spherical spinors. Inserting Eq.
(\ref{opera}) into Eq. (\ref{field}) we obtain the following
equations for the upper component of the nucleon's  wave function
 \begin{equation}
E_{\alpha}G_{\alpha}(r) = \left[- \frac{d}{dr} + W(r)
 \right] M_{eff}^{-1} \left[ \frac{d}{dr} + W(r) \right]
 G_{\alpha}(r) + U_{eff}G_{\alpha}(r),  \label{schn}
 \end{equation}
and the lower component of the anti-nucleon's wave function
 \begin{equation}
\bar{E}_{\alpha}\bar{G}_{\alpha}(r) = \left[-
\frac{d}{dr}+\bar{W}(r) \right] \bar{M}_{eff}^{-1} \left[
\frac{d}{dr} + \bar{W}(r) \right]
 \bar{G}_{\alpha}(r) + \bar{U}_{eff}\bar{G}_{\alpha}(r).
 \label{schan}
 \end{equation}
Other components can be obtained through the following relations
 \begin{eqnarray}
&& F_{\alpha}(r) = M_{eff}^{-1} \left[ \frac{d}{dr} + W(r)
   \right] G_{\alpha}(r), \\
&& \bar{F}_{\alpha}(r) = \bar{M}_{eff}^{-1} \left[ \frac{d}{dr}
   + \bar{W}(r)\right] \bar{G}_{\alpha}(r).
 \end{eqnarray}
The Schr\"{o}dinger-equivalent effective masses and potentials are
defined as follows: for the nucleon
 \begin{eqnarray}
&&  M_{eff}= E_{\alpha}+ M_{N} - {\rm g}_{\sigma}\sigma (r)
 - {\rm g}_{\omega}\omega_{0}(r) -\frac{1}{2}{\rm
 g}_{\rho}\tau_{0\alpha}R_{0,0}(r) -\frac{1}{2}e
 \left(1+\tau_{0\alpha}\right) A_{0}(r), \label{effmn} \\
&& U_{eff}= M_{N} - {\rm g}_{\sigma}\sigma(r)
 + {\rm g}_{\omega}\omega_{0}(r)+\frac{1}{2}{\rm
 g}_{\rho}\tau_{0\alpha}R_{0,0}(r) + \frac{1}{2}e\left(
 1+\tau_{0\alpha}\right) A_{0}(r), \\
 && W(r)=\frac{\kappa_{\alpha}}{r}-\frac{f_{\omega}}{2M_{N}}\left(
    \partial_{r}\omega_{0}(r)\right)
    -\frac{f_{\rho}}{4M_{N}}\tau_{0\alpha}\left(\partial_{r}R_{0,0}(r)\right),
 \end{eqnarray}
for the anti-nucleon
 \begin{eqnarray}
&&  \bar{M}_{eff}= \bar{E}_{\alpha}+
    M_{N}-{\rm g}_{\sigma}\sigma(r)
 + {\rm g}_{\omega}\omega_{0}(r) -\frac{1}{2}{\rm
 g}_{\rho}\tau_{0\alpha}R_{0,0}(r) +\frac{1}{2}e
 \left(1+\tau_{0\alpha}\right) A_{0}(r), \\
&& \bar{U}_{eff}= M_{N} - {\rm g}_{\sigma}\sigma(r)
 - {\rm g}_{\omega}\omega_{0}(r)+\frac{1}{2}{\rm
 g}_{\rho}\tau_{0\alpha}R_{0,0}(r) - \frac{1}{2}e\left(
 1+\tau_{0\alpha}\right) A_{0}(r), \\
 && \bar{W}(r)=\frac{\kappa_{\alpha}}{r}+\frac{f_{\omega}}{2M_{N}}\left(
    \partial_{r}\omega_{0}(r)\right)
    -\frac{f_{\rho}}{4M_{N}}\tau_{0\alpha}\left(\partial_{r}R_{0,0}(r)\right).
  \label{wan}
 \end{eqnarray}
One can see that the difference between the equations of nucleons
and anti-nucleons  relies only on the definition of the effective
masses and potentials.  The main ingredients are meson fields
which can be obtained through solving the Laplace equations of
mesons. The source terms  are  various densities containing the
contributions both from the valence nucleons and the Dirac sea.
They are evaluated by means of the derivative expansion technique
\cite{Ait84}. The energy spectra of the nucleon and the
anti-nucleon are computed by means of the  equations
\begin{eqnarray}
 E_{\alpha} &=& \int^{\infty}_{0} dr \lbrace G_{\alpha}(r) \left[
 -\frac{d}{dr} + W(r) \right] F_{\alpha}(r) + F_{\alpha}(r)
 \left[ \frac{d}{dr} + W(r) \right] G_{\alpha}(r)
 \nonumber \\
&&  + G_{\alpha}(r)U_{eff}G_{\alpha}(r) - F_{\alpha}(r) \left[
 M_{eff}- E_{\alpha} \right] F_{\alpha}(r) \rbrace, \\
 \vspace{0.4cm}
 \bar{E}_{\alpha} &=& \int^{\infty}_{0} dr \lbrace
\bar{G}_{\alpha}(r)\left[ -\frac{d}{dr} + \bar{W}(r) \right]
\bar{F}_{\alpha}(r)
 + \bar{F}_{\alpha}(r)
 \left[\frac{d}{dr} + \bar{W}(r) \right]\bar{G}_{\alpha}(r)
  \nonumber \\
&&  + \bar{G}_{\alpha}(r)\bar{U}_{eff}\bar{G}_{\alpha}(r)
 - \bar{F}_{\alpha}(r) \left[ \bar{M}_{eff}
 - \bar{E}_{\alpha} \right] \bar{F}_{\alpha}(r) \rbrace.
 \end{eqnarray}

The parameters of the model are fitted to the properties of
spherical nuclei \cite{Mao99,Mao03}. The major result is that a
large effective nucleon mass $m^{*}/M_{N} \approx 0.78$ is
obtained. In Table 1 we present the single-particle energies of
protons (neutrons) and anti-protons (anti-neutrons) in
 $^{16}{\rm O}$, $^{40}{\rm Ca}$ and $^{208}{\rm Pb}$.
 The binding energies per nucleon and the {\em
rms} charge radii are given too. The experimental data are taken
from Ref. \cite{Mat65}. It can be found that  the relativistic
Hartree approach taking into account the vacuum effects can
reproduce the observed binding energies , {\em rms} charge radii
and particle spectra quite well. Because of the large effective
nucleon mass, the spin-orbit splitting on the $1p$ levels is
rather small in the RHA1 model. The situation has been ameliorated
conspicuously in the RHAT model incorporating the tensor couplings
for the $\omega$- and $\rho$-meson, while a large $m^{*}$ stays
unchanged. On the other hand, the anti-particle energies
calculated with the RHAT set of parameters are 20 -- 30 MeV larger
than that reckoned with the RHA1 set.

\begin{table}
Table 1. The single-particle energies of both protons (neutrons)
and anti-protons (anti-neutrons) as well as the binding energies
per nucleon and the {\em rms} charge radii in $^{16}{\rm O}$,
$^{40}{\rm Ca}$ and
 $^{208}{\rm Pb}$.\\
{\small
\begin{tabular}{cccccccc}
\hline
 & RHA1 & RHAT & Expt. & & RHA1 & RHAT & Expt.\\
 \hline
$\;\;\;\;\;\;\;\;^{16}{\rm O}$  &        &      &  &&&&     \\
$E/A$ (MeV)      &   8.00  & 7.94 &  7.98 &&&&\\
$r_{ch}$ (fm)    &   2.66  & 2.64 &  2.74 &&&&\\
 PROTONS         &         &      &       &
 NEUTRONS        &         &      &       \\
$1s_{1/2}$ (MeV) &   30.68 & 31.63&  40$\pm$8 &
$1s_{1/2}$ (MeV) &  34.71   & 35.78 &  45.7 \\
$1p_{3/2}$ (MeV) &   15.23 & 16.18&  18.4  &
$1p_{3/2}$ (MeV)  &  19.04   & 20.18 &  21.8 \\
$1p_{1/2}$ (MeV) &   13.24 & 12.22&  12.1  &
$1p_{1/2}$ (MeV)  &  17.05   & 15.75 &  15.7 \\
 ANTI-PRO.       &         &      &          &
 ANTI-NEU.       &         &       &          \\
$1\bar{s}_{1/2}$ (MeV) &   299.42 & 328.55 &   &
$1\bar{s}_{1/2}$ (MeV) & 293.23  & 322.47&     \\
$1\bar{p}_{3/2}$ (MeV) &   258.40 & 283.44 &  &
$1\bar{p}_{3/2}$ (MeV)& 252.48  & 277.94&      \\
$1\bar{p}_{1/2}$ (MeV) &   258.93 & 285.87 &   &
$1\bar{p}_{1/2}$ (MeV)& 252.97  & 279.22&      \\
\hline
$\;\;\;\;\;\;\;^{40}{\rm Ca}$ &       &      &     &&&&  \\
$E/A$ (MeV)      &   8.73   &  8.62   &  8.55  &&&& \\
$r_{ch}$ (fm)    &   3.42   &  3.41   &  3.45  &&&& \\
PROTONS          &          &         &      &
NEUTRONS         &          &         &      \\
$1s_{1/2}$ (MeV) &   36.58  &  37.01  & 50$\pm$11 &
$1s_{1/2}$ (MeV)  &  44.48   & 44.98 &     \\
$1p_{3/2}$ (MeV) &   25.32  &  25.95  &     &
$1p_{3/2}$ (MeV)  &  32.98   & 33.83 &       \\
$1p_{1/2}$ (MeV) &   24.03  &  23.63  &34$\pm$6 &
$1p_{1/2}$ (MeV)  &  31.71   & 30.99 &  \\
ANTI-PRO.        &             &         &     &
ANTI-NEU.         &          &       &              \\
$1\bar{s}_{1/2}$ (MeV)& 339.83 &  367.90 &    &
$1\bar{s}_{1/2}$ (MeV)& 327.96  & 355.70&      \\
$1\bar{p}_{3/2}$ (MeV)& 309.24 &  332.10 &     &
$1\bar{p}_{3/2}$ (MeV)& 298.04  & 321.07&       \\
$1\bar{p}_{1/2}$ (MeV)& 309.52 &  333.37 &     &
$1\bar{p}_{1/2}$ (MeV)& 298.26  & 322.15&          \\
\hline
$\;\;\;\;\;\;\;^{208}{\rm Pb}$ &        &      &       \\
$E/A$ (MeV)      &   7.93   &  7.88   &  7.87   \\
$r_{ch}$ (fm)    &   5.49   &  5.46   &  5.50   \\
 PROTONS         &         &         &     &
 NEUTRONS        &           &       &         \\
$1s_{1/2}$ (MeV) &   40.80  &  41.74  &     &
$1s_{1/2}$ (MeV)  &   47.40   & 46.70 &      \\
$1p_{3/2}$ (MeV) &    36.45  &  37.38  &    &
$1p_{3/2}$ (MeV)  &   42.66   & 42.31 &          \\
$1p_{1/2}$ (MeV) &    36.21  &  37.18  &    &
$1p_{1/2}$ (MeV)  &   42.45   & 41.64 &            \\
 ANTI-PRO.       &           &         &     &
 ANTI-NEU.         &           &       &         \\
$1\bar{s}_{1/2}$ (MeV)&354.18  &  377.37 &    &
$1\bar{s}_{1/2}$ (MeV)& 313.18  & 334.39&         \\
$1\bar{p}_{3/2}$ (MeV)&344.48  &  366.95 &    &
$1\bar{p}_{3/2}$ (MeV)& 304.61  & 325.41&       \\
$1\bar{p}_{1/2}$ (MeV)&344.52  &  367.24 &    &
$1\bar{p}_{1/2}$ (MeV)& 304.61  & 325.28&         \\
\hline
\end{tabular}
 }
\end{table}
\vspace{-0.2cm}
\section*{Acknowledgements}
The authors thank P.-G.~Reinhard, Z.Y.~Zhang and Z.Q.~Ma for
stimulating discussions. This work was supported by the National
Natural Science Foundation of China under Grant No. 10275072.

\end{document}